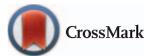

# Proper time operator and its uncertainty relation

Hou Y Yau 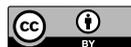

San Francisco State University, 1600 Holloway Avenue, San Francisco, CA, United States of America

E-mail: hyau@mail.sfsu.edu

**Keywords:** time and space symmetry, uncertainty relation, time operator, vibration in time, bosonic field, Klein–Gordon equation

## Abstract

We study the quantum properties of an oscillator in proper time. This proper time oscillator is a particle model with mass that is on shell. Its internal time can be treated as a self-adjoint operator. The displaced time and displaced time rate of the oscillator obey an uncertainty relation resembling the one between position and momentum, which is different from the usual energy-time uncertainty relation. In addition, we demonstrate that a matter field with proper time oscillators satisfies the Klein–Gordon equation. It has the properties of a zero-spin quantum field. The formulations adopted permit a more symmetrical treatment between time and space in a matter field.

## 1. Introduction

In quantum theory, physical quantities such as position and momentum are represented by operators. This principle plays a vital role in the formulation of the theory. However, when we talk about time, the situation is very different. Time, in general, is postulated as a parameter in quantum theory. It cannot be treated as an operator based on an argument propounded by Pauli [1, 2]. Contrary to what we have learned from general relativity, there is nothing dynamic about time in quantum physics.

According to Pauli's theorem, time and energy are a conjugate pair that shall satisfy a commutation relation $[H, t] = -i$. However, the spectrum of a Hamiltonian is either discrete or bounded from below. Consequently, time cannot be taken as a self-adjoint operator with a spectrum that spans the whole real line. One of the widely accepted ways to overcome this conundrum is to invoke the use of positive operator valued measures (POVM) as solutions [3–9]. Besides these efforts, other models and methods have been engendered to allow a more dynamic treatment of time in the classical and quantum theories [10–13]. The asymmetric treatment of time and space in quantum theory has inspired the quest for a time operator [14–23] to restore symmetry in its formulations.

The constraint that time can only be treated as a parameter has other fundamental implications. In Heisenberg's uncertainty principle, a particle's position and momentum are dynamic variables that can vary over time. On the contrary, time itself is an independent variable on which everything else evolves against. Evidently, the energy-time uncertainty principle must hold a different type of relationship. The disparity of how space and time are handled has led to many different ways to interpret the energy-time uncertainty principle; each is valid in its own contexts [24–27]. The most common ones are interpreted as a constraint on the state preparation, or on the statistical errors and disturbances [28–31].

To restore the symmetry in a matter field, we can allow the matter to vibrate in both space and time [32–36]. In this paper, we demonstrate that a proper time oscillating system can be defined in a similar way as we formulate a one-dimensional quantum harmonic oscillator. Its concepts can be applied to a matter field. A system with vibrations of matter in time has properties that resemble those for a spin-zero quantum field. The system obeys the Klein–Gordon equation, which can be treated as a quantized field. Besides, the particle observed has oscillation in proper time. Its energy is on shell. The internal time of the proper time oscillator can be reckoned as a self-adjoint operator. Its displaced time forms a conjugate pair with the displaced time rate. Unlike the typical energy-time uncertainty principle, this displaced time-displaced time rate conjugate pair satisfies an uncertainty relation that resembles the one between position and momentum. Not only the proper time oscillator can reconcile the properties of a quantum particle, but its formulations also permit a more symmetrical treatment of time and space. The oscillation of matter in time is an additional degree of freedom introduced for restoring the symmetry between time and space in a matter field.

The paper is organized as follows: section 2 elaborates on a proper time oscillator's basic properties. In section 3, we introduce a Lorentz covariant plane wave with vibrations of matter in time and space. In sections 4 and 5, we study the quantized proper time oscillator at rest and under motion. In section 6, we compare a quantum harmonic oscillator's properties with a field that has vibration of matter in proper time. In section 7, we study the quantum properties of the field with vibrations of matter in space and time. A self-adjoint internal time operator for the system is introduced. The last section of this paper is reserved for conclusions and discussions.

## 2. Proper time oscillator

Let us consider the background coordinates $(t, \mathbf{x})$ of a flat spacetime as observed in an inertial frame $O$. Time in this background is the 'external time' as measured by a clock at spatial infinity that is not coupled to the system under investigation [28, 29, 37, 38]. This external time $t$ is a parameter and an independent variable in the equation of motion. There is nothing dynamic about its nature. We will use the background coordinates as references to measure the oscillations of matter that we will study.

At the spatial origin $\mathbf{x}_0$ of the coordinate system, let us introduce a point mass $m$ that we will allow to oscillate in proper time. The equation of motion for this oscillation can be defined as,

$$t_f = t + t_d = t - T_0 \sin(\omega_0 t), \qquad (1)$$

where

$$t_d = -T_0 \sin(\omega_0 t). \qquad (2)$$

This equation is analogous to the one defined for a classical simple harmonic oscillator, except we are considering the oscillation in the temporal direction. Furthermore, this oscillator is stationary in space. There is no oscillation in the spatial direction.

The internal time $t_f$ is a dynamic variable of the system, which is a function of $t$. Time measured by an internal clock of the oscillator will oscillate with varying rates relative to the external time,

$$u_f = \frac{\partial t_f}{\partial t} = 1 - T_0 \omega_0 \cos(\omega_0 t). \qquad (3)$$

The average of this internal time rate is 1. If the oscillation is rapid, the oscillator will appear to travel along a time-like geodesic when averaged over time. Analogous to the classical system, we will define the amplitude $T_0$ as the maximum time displaced between the internal time $t_f$ and the external time $t$.

The displaced time $t_d$ is measured relative to the external time $t$. Time measured by an internal clock of the point mass is displaced from the external time as it propagates. The equilibrium position of this oscillator is the moving external time. Relative to this equilibrium, we can also define an oscillating displaced time rate,

$$u_d = \frac{\partial t_d}{\partial t} = -T_0 \omega_0 \cos(\omega_0 t). \qquad (4)$$

The oscillating displaced time $t_d$ is analogous to the spatial displacement observed in a classical oscillator. As the internal time is displaced from its equilibrium position, the oscillator will try to restore its equilibrium. After the internal time reaches its equilibrium, the non-zero oscillating displaced time rate $u_d$ will cause the internal time to pass over the equilibrium.

As we have learned from wave mechanics, a simple harmonic oscillator's energy is proportional to the square of its amplitude. We expect the proper time oscillator will hold a similar relationship. Since the oscillator is stationary in space, there is no associated kinetic energy. The energy of this oscillator shall correspond to certain intrinsic energy of the point mass. However, the system that we are studying involves only time oscillation of matter with no other energy or force fields. The sole source of energy is the internal mass-energy of the point mass. Here, we will consider the energy of the proper time oscillator as internal mass-energy.

Taking the proper time oscillator as a particle, the energy of this oscillator must be on shell. Only oscillators with the specific mass-energy of the particle can be observed. For a system with multiple numbers of proper time oscillators, the total energy is the sum of the internal mass-energy of all the particles present, which is quantized. Based on this property, we will show that a proper time oscillating system can be quantized in a similar way as we quantize a quantum harmonic oscillator. The results are similar, except the spatial oscillation is replaced by the temporal oscillation. However, there is one fundamental difference. Instead of dealing with the different quantized energy levels of a quantum harmonic oscillator, the proper time oscillating system shall be treated as a field that allows the creation and annihilation of particles.

## 3. Lorentz covariant plane wave

In the previous section, we have considered the oscillation of a point mass in proper time. Here, we will apply the same concept to a matter field. However, before we proceed, we shall be reminded that a proper time displacement is only the 0-component of a 4-displacement. In a matter field, we shall consider the relative displacements of matter in both the temporal and spatial directions, i.e.,

$$t_f = t + T\sin(\mathbf{k}\cdot\mathbf{x} - \omega t) = t + t_d = t + \mathrm{Re}(\zeta_t), \tag{5}$$

$$\mathbf{x}_f = \mathbf{x} + \mathbf{X}\sin(\mathbf{k}\cdot\mathbf{x} - \omega t) = \mathbf{x} + \mathbf{x}_d = \mathbf{x} + \mathrm{Re}(\boldsymbol{\zeta}_x), \tag{6}$$

where

$$t_d = \mathrm{Re}(\zeta_t) = T\sin(\mathbf{k}\cdot\mathbf{x} - \omega t), \tag{7}$$

$$\mathbf{x}_d = \mathrm{Re}(\boldsymbol{\zeta}_x) = \mathbf{X}\sin(\mathbf{k}\cdot\mathbf{x} - \omega t), \tag{8}$$

$$\zeta_t = -iTe^{i(\mathbf{k}\cdot\mathbf{x} - \omega t)}, \quad \boldsymbol{\zeta}_x = -i\mathbf{X}e^{i(\mathbf{k}\cdot\mathbf{x} - \omega t)}, \tag{9}$$

$$T = (\omega/\omega_0)T_0, \quad \mathbf{X} = (\mathbf{k}/\omega_0)T_0. \tag{10}$$

In the following analysis, we will begin by treating the wave as a classical system, and then quantize it as a quantum field.

The amplitude $(T, \mathbf{X})$ is a Lorentz transformation of a proper time amplitude $T_0$, i.e., $(T_0, 0, 0, 0) \to (T, \mathbf{X})$, where $T^2 = T_0^2 + |\mathbf{X}|^2$. Analogous to a classical wave, we will define the temporal amplitude $T$ as the maximum difference between the internal time $t_f$ of the matter's clock and the external time $t$. Similarly, the spatial amplitude $\mathbf{X}$ is the maximum displacement of matter in the wave from its equilibrium coordinate $\mathbf{x}$.

The internal time of matter inside this wave is propagating at varying time rates. Apart from the temporal vibration, matter also vibrates in spatial directions. To study these vibrations, we will adopt a convention similar to the Lagrangian formulation in wave mechanics. The spatial displacement $\mathbf{x}_d$ is defined as the spatial measurement relative to the undisturbed state labeled $\mathbf{x}$. Similarly, the displaced time $t_d$ is defined as the time difference relative to the external time $t$. Therefore, $\mathbf{x}_d(t, \mathbf{x})$ is the displacement of matter at time $t$, which has coordinate $\mathbf{x}$ in the undisturbed condition. The displaced time $t_d(t, \mathbf{x})$ is the difference in time, measured originally from $\mathbf{x}$, relative to the coordinate time $t$. Consequently, matter originally at $\mathbf{x}$ will be displaced to $\mathbf{x}_f = \mathbf{x} + \mathbf{x}_d$, with an internal time $t_f = t + t_d$, instead of the time $t$ at spatial infinity.

The vibrations $\zeta_t$ and $\zeta_\mathbf{x}$ defined in equation (9) can be described by a Lorentz covariant plane wave,

$$\zeta = \frac{T_0}{\omega_0}e^{i(\mathbf{k}\cdot\mathbf{x} - \omega t)}, \tag{11}$$

such that

$$\zeta_t = \partial_0\zeta, \quad \boldsymbol{\zeta}_\mathbf{x} = -\boldsymbol{\nabla}\zeta. \tag{12}$$

$T_0$, $T$, and $\mathbf{X}$ are to be treated as complex amplitudes. As we shall note, the Lorentz covariant plane wave $\zeta$ and its complex conjugate $\zeta^*$ satisfy the wave equations:

$$\partial_u\partial^u\zeta + \omega_0^2\zeta = 0, \tag{13}$$

$$\partial_u\partial^u\zeta^* + \omega_0^2\zeta^* = 0. \tag{14}$$

Equations (13) and (14) are similar to the Klein–Gordon equation. The corresponding Lagrangian density for the equations of motion is,

$$\mathcal{L} = K[(\partial^u\zeta^*)(\partial_u\zeta) - \omega_0^2\zeta^*\zeta], \tag{15}$$

and the Hamiltonian density is

$$\mathcal{H} = K[(\partial_0\zeta^*)(\partial_0\zeta) + (\boldsymbol{\nabla}\zeta^*)\cdot(\boldsymbol{\nabla}\zeta) + \omega_0^2\zeta^*\zeta], \tag{16}$$

where $K$ is a constant of the system. Substitute equation (11) into equation (16), the Hamiltonian density of a plane wave is

$$\mathcal{H} = K(T^*T + \mathbf{X}^*\cdot\mathbf{X} + T_0^*T_0), \tag{17}$$

which is the summation of the energies for the oscillations of matter in time and space. Each term in the summation is the square of an oscillating amplitude.

## 4. Oscillator at rest

To study the properties of the Lorentz covariant waves, let us first consider a plane wave with vibrations in proper time only ($\omega = \omega_0$, $|\mathbf{k}| = 0$), i.e.,

$$\zeta_0 = \frac{T_0}{\omega_0} e^{-i\omega_0 t}. \tag{18}$$

From equation (17), the Hamiltonian density is $\mathcal{H}_0 = 2K T_0^* T_0$. This result is similar to the Hamiltonian density of a classical harmonic oscillating system, except that the vibration is in time. For a system in a box with volume $V$ that can have multiple numbers of particles with mass $m$, we make an ansatz[1],

$$K = \frac{m\omega_0^2}{2V}. \tag{19}$$

The Hamiltonian density for the proper time plane wave becomes,

$$\mathcal{H}_0 = \frac{m\omega_0^2 T_0 T_0^*}{V}. \tag{20}$$

The Hamiltonian density $\mathcal{H}_0$ corresponds to an energy $E = m\omega_0^2 T_0 T_0^*$ in volume $V$. On the other hand, the plane wave $\zeta_0$ involves only the temporal oscillation of particles with no other force field or energy. The sole source of energy in this system is the mass-energy of the particles, i.e.

$$E = nm = m\omega_0^2 T_0 T_0^*, \tag{21}$$

and

$$n = \omega_0^2 T_0 T_0^*, \tag{22}$$

is the number of particles. Since the energy of a particle must be on shell, energy $E$ is quantized.

In a normalized plane wave $\zeta_0$, equation (22) becomes

$$\omega_0^2 T_0 T_0^* = 1. \tag{23}$$

This implies that a particle can only be observed with a specific proper time oscillation amplitude, $|\mathring{T}_0| = 1/\omega_0$. From equation (1), the internal time $\mathring{t}_f$ of the proper time oscillator can be written as

$$\mathring{t}_f = t - \frac{\sin(\omega_0 t)}{\omega_0}. \tag{24}$$

The rate of this internal time relative to the external time is

$$\frac{\partial \mathring{t}_f}{\partial t} = 1 - \cos(\omega_0 t). \tag{25}$$

Not only the average of this time rate is 1, but its value is bounded between 0 and 2, which is positive. Therefore, the internal time of the oscillator cannot go back to its past. It moves only forward relative to the external time.

In the system considered, the mass-energy of a particle is generated by its oscillation in proper time. The particles observed in the plane wave $\zeta_0$ are at rest with their own de Broglie's internal clock, i.e. $\omega_0 = m$. Unlike time dilation in relativity, the oscillation of matter in time is not the result of relative movement or gravity; it is an additional degree of freedom introduced for restoring the symmetry between time and space in a matter field.

Taking the proper time oscillator as a particle with the mass of an electron ($\omega_0 = 7.6 \times 10^{20} s^{-1}$ and $\mathring{T}_0 = 1/\omega_0 = 1.32 \times 10^{-21} s$), the amplitude of the oscillation is small, but the frequency is high. If the observer's clock is not sensitive enough to detect the high frequency of vibration [39], the particle will appear to travel along a time-like geodesic. Despite that, the effects of the temporal oscillation are theoretically observable. For example, the decay rates of an unstable particle will vary at different phases of a cycle.

## 5. Moving oscillator

In this section, let us consider another plane wave,

$$\tilde{\zeta} = \sqrt{\frac{\omega_0}{\omega}}\zeta = \frac{T_0}{\sqrt{\omega\omega_0}} e^{i(\mathbf{k}\cdot\mathbf{x} - \omega t)}, \tag{26}$$

where the Lorentz covariant plane wave $\zeta$ is defined in equation (11), and $\sqrt{\omega_0/\omega}$ is a normalization factor. Based on equations (5) and (6), matters in this plane wave have vibrations in both the temporal and spatial

---

[1] In equation (17), the second term on the right-hand side of the Hamiltonian density equation is $\mathcal{H}_x = K\mathbf{X}^* \cdot \mathbf{X}$. In the non-relativistic limit, our choice of $K = m\omega_0^2/2V$ gives us a familiar expression for the Hamiltonian density of a system with oscillation in space.

directions,

$$\tilde{t}_f = t + \text{Re}(\tilde{\zeta}_t), \tag{27}$$

$$\tilde{\mathbf{x}}_f = \mathbf{x} + \text{Re}(\tilde{\zeta}_\mathbf{x}), \tag{28}$$

where

$$\tilde{\zeta}_t = \partial_0 \tilde{\zeta} = -i\tilde{T} e^{i(\mathbf{k}\cdot\mathbf{x}-\omega t)}, \tag{29}$$

$$\tilde{\zeta}_\mathbf{x} = -\boldsymbol{\nabla}\tilde{\zeta} = -i\tilde{\mathbf{X}} e^{i(\mathbf{k}\cdot\mathbf{x}-\omega t)}, \tag{30}$$

$$\tilde{T} = \sqrt{\frac{\omega_0}{\omega}} T = T_0 \sqrt{\frac{\omega}{\omega_0}}, \quad \tilde{\mathbf{X}} = \sqrt{\frac{\omega_0}{\omega}} \mathbf{X} = \frac{T_0 \mathbf{k}}{\sqrt{\omega_0 \omega}}, \tag{31}$$

where $T$ and $\mathbf{X}$ are the amplitudes of the Lorentz covariant plane wave $\zeta$ defined in equation (10). Therefore, the amplitudes $\tilde{T}$ and $\tilde{\mathbf{X}}$ of the normalized plane wave $\tilde{\zeta}$ is the Lorentz transformation of the proper time oscillation amplitude multiplied by a normalization factor $\sqrt{\omega_0/\omega}$.

From equation (16), the Hamiltonian density of $\tilde{\zeta}$ is,

$$\tilde{\mathcal{H}} = \frac{m\omega\omega_0 T_0^* T_0}{V}. \tag{32}$$

With $T_0 = 1/\omega_0$, the Hamiltonian density of a normalized plane wave is $\tilde{\mathcal{H}} = \omega/V$, which corresponds to an oscillator with energy $\omega$ in volume $V$. The particle observed is traveling with a velocity $\mathbf{v} = \mathbf{k}/\omega$. The temporal and spatial vibrations of the particle that follows the path $\mathbf{x} = \mathbf{v}t$ in the plane wave $\tilde{\zeta}$ are,

$$\mathring{t}_f = t - \mathring{T} \sin(\omega_p t), \tag{33}$$

$$\mathring{\mathbf{x}}_f = \mathbf{v}t - \mathring{\mathbf{X}} \sin(\omega_p t), \tag{34}$$

where

$$\mathring{T} = \sqrt{\frac{\omega}{\omega_0^3}}, \quad \mathring{\mathbf{X}} = \frac{\mathbf{k}}{\sqrt{\omega_0^3 \omega}} = \sqrt{\frac{\omega}{\omega_0^3}} \mathbf{v}, \quad \omega_p = \frac{\omega_0^2}{\omega}. \tag{35}$$

The particle is assumed to be first observed at the origin of the spatial coordinates $\mathbf{x}_0$ at $t = 0$. The internal time rate relative to the external time is,

$$\frac{\partial \mathring{t}_f}{\partial t} = 1 - \sqrt{\frac{\omega_0}{\omega}} \cos(\omega_p t). \tag{36}$$

The observed velocity with oscillation is,

$$\frac{\partial \mathring{\mathbf{x}}_f}{\partial t} = \mathbf{v} \left[ 1 - \sqrt{\frac{\omega_0}{\omega}} \cos(\omega_p t) \right]. \tag{37}$$

As shown, a particle has oscillations in the temporal and spatial directions when it is in motion. From equation (35), the oscillator will have a lower frequency and larger amplitudes of oscillation at a higher speed. Consequently, it will be easier to detect the effects of the oscillations at higher energy. For example, it could be possible to detect the spatial oscillations associated with the temporal oscillation. In particular, two oscillators with the same average velocity can reach a target at slightly different times, depending on the relative phase of their oscillations. In theory, this deviation can be observable by repeated measurements of the oscillators' arrival times at a detector as referenced to a clock in the laboratory that is not coupled to the system under investigation.

To demonstrate the magnitude of the oscillations, let us consider the oscillator as a neutrino with an assuming mass of $m = 2$ eV [40, 41]. From equation (35),

$$E = 1 \text{ Gev} \quad \Rightarrow \quad \mathring{T} = 7.4 \times 10^{-12} \text{ s}, \quad |\mathring{\mathbf{X}}| = 0.22 \text{ cm}, \quad \omega_p = 6.1 \times 10^6 \text{ s}^{-1}. \tag{38}$$

These quantities are not outrageously small/large for a neutrino moving at near light speed. Because of their extremely light weight, neutrinos can be projected to a very high speed, which can amplify the oscillations for measurements.

## 6. Field with proper time oscillations

The plane wave $\zeta_0$ and its conjugate $\zeta_0^*$ can be superposed to form a field with oscillations of matter in proper time, i.e.

$$\zeta' = \frac{1}{\sqrt{2}}[\zeta_0 + \zeta_0^*] = \frac{1}{\sqrt{2}\,\omega_0}[T_0 e^{-i\omega_0 t} + T_0^* e^{i\omega_0 t}]. \tag{39}$$

As discussed, the particles observed in this field are at rest and can only oscillate with a specific amplitude $|T_0| = 1/\omega_0$. The system is quantized, and the proper time oscillators are the field quanta. As we have learned from quantum theory, the transition of a classical field to a quantum field can be done via canonical quantization. Quantities such as $\zeta'$ and $T_0$ can be promoted to operators.

Analogous to a one-dimensional quantum harmonic oscillator, we can define a creation operator $a^\dagger$ and an annihilation operator $a$, and promote the number of particles $n$ from equation (22) to an operator, i.e.

$$a^\dagger = \omega_0 T_0^\dagger, \tag{40}$$

$$a = \omega_0 T_0, \tag{41}$$

$$N = a^\dagger a = \omega_0^2 T_0^\dagger T_0. \tag{42}$$

The operators $a$ and $a^\dagger$ satisfy the commutation relation,

$$[a, a^\dagger] = 1. \tag{43}$$

Substitute $\zeta'$ from equation (39) into equation (16), and integrate over the volume $V$, the total Hamiltonian of the system is

$$H' = \omega_0 \left(a^\dagger a + \frac{1}{2}\right), \tag{44}$$

where we have replaced mass $m$ with the de Broglie's angular frequency $\omega_0$ [42].

From equation (12), the displaced time $t'_d$ and the displaced time rate $u'_d$ in the field $\zeta'$ are,

$$t'_d = \zeta'_t = \partial_0 \zeta' = \frac{-i}{\sqrt{2}}[T_0 e^{-i\omega_0 t} - T_0^\dagger e^{i\omega_0 t}] = \frac{-i}{\sqrt{2}\,\omega_0}[a e^{-i\omega_0 t} - a^\dagger e^{i\omega_0 t}], \tag{45}$$

$$u'_d = \partial_0 t'_d = \frac{-\omega_0}{\sqrt{2}}[T_0 e^{-i\omega_0 t} + T_0^\dagger e^{i\omega_0 t}] = \frac{-1}{\sqrt{2}}[a e^{-i\omega_0 t} + a^\dagger e^{i\omega_0 t}], \tag{46}$$

where both $t'_d$ and $u'_d$ are real. The Hamiltonian from equation (44) can be rewritten as

$$H' = \frac{m}{2}(\omega_0^2 t'^2_d + u'^2_d). \tag{47}$$

Despite the oscillation is in time, equation (47) is the familiar expression for a harmonic oscillator's Hamiltonian.

From equations (43), (45) and (46), the displaced time and the displaced time rate shall satisfy a commutation relation,

$$[t'_d, m u'_d] = i. \tag{48}$$

Consequently, $t'_d$ and $u'_d$ are a conjugate pair. Both of them are self-adjoint operators. As we shall note, time oscillation can be displaced either in the positive or negative direction relative to the equilibrium position. The spectrum of the displaced time can span the whole real line. Furthermore, $m u'_d$ plays a similar role as 'momentum' in the temporal oscillation. However, it is not the Hamiltonian of the system. The displaced time $t'_d$ and the Hamiltonian $H'$ do not form a conjugate pair. Therefore, there is no restriction that the spectrum of $t'_d$ has to be bounded, as stipulated by Pauli's theorem.

It is interesting to note that $\zeta'$ and $t'_d$ can also form a conjugate pair, which satisfy an uncertainty relation

$$[\zeta', t'_d] = i \omega_0^{-3}. \tag{49}$$

The choice of using either $\zeta'$ or $u'_d$ is just a matter of convenience. Both ways will lead us to the same conclusion that $t'_d$ does not form a conjugate pair with the Hamiltonian $H'$. The operators $\zeta'$, $t'_d$ and $u'_d$ are Hermitian.

Based on equation (5), the internal time $t'_f$ is the summation of the external time $t$ and the displaced time $t'_d$, i.e., $t'_f = t + t'_d$. However, the external time $t$ is a parameter, and the displaced time $t'_d$ is a self-adjoint operator. Their summation must also be a self-adjoint operator. The result is that the internal time $t'_f$ and the displaced time rate $u'_d$ shall satisfy a similar commutation relation as equation (48), i.e.,

$$[t'_f, m u'_d] = i. \tag{50}$$

Again, $t'_f$ does not form a conjugate pair with the Hamiltonian $H'$. There is no restriction that $t'_f$ cannot be treated as a self-adjoint operator.

Using the standard properties of the creation and annihilation operators, we can obtain the variance for the displaced time and the displaced time rate,

Table 1. Comparing the proper time oscillator with the quantum harmonic oscillator

|  | Proper time oscillator | Quantum harmonic oscillator |
|---|---|---|
| Hamiltonian | $H' = \omega_0(a^\dagger a + \frac{1}{2})$ | $H = \omega(a^\dagger a + \frac{1}{2})$ |
| Commutation Relation | $[t_d', mu_d'] = i$ | $[x,p]=i$ |
| Uncertainty Relation | $\Delta t_d' \Delta(mu_d') \geq \frac{1}{2}$ | $\Delta x \Delta p \geq \frac{1}{2}$ |

$$\Delta t_d' = \frac{1}{\omega_0}\sqrt{\left(n + \frac{1}{2}\right)}, \quad (51)$$

$$\Delta u_d' = \sqrt{n + \frac{1}{2}}, \quad (52)$$

which lead to an uncertainty relation,

$$\Delta t_d' \Delta(mu_d') = n + \frac{1}{2} \geq \frac{1}{2}. \quad (53)$$

The displaced time operator is an analogy of the position operator. It satisfies an uncertainty relation with the displaced time rate that resembles the one between position and momentum. Unlike the problem with the standard time-energy uncertainty relation, the displaced time $t_d'$ and the internal time $t_f'$ can be treated as self-adjoint operators with spectra that can span the whole real line.

As shown in table 1, the field $\zeta'$ with proper time oscillations has resembling properties as the quantum harmonic oscillator, except the oscillation is in time. Both systems have similar Hamiltonian, commutation relation, and uncertainty relation. The concept of a quantum harmonic oscillator can be extended to a matter field by considering the oscillation of matter in time.

## 7. Matter Field

The plane waves $\tilde{\zeta}$ and their conjugates $\tilde{\zeta}^*$ can be superposed to form a field with vibrations of matter in space and time,

$$\zeta(x) = \frac{1}{\sqrt{2}}\sum_{\mathbf{k}}[\tilde{\zeta}_{\mathbf{k}}(x) + \tilde{\zeta}_{\mathbf{k}}^*(x)], \quad (54)$$

where periodic boundary conditions are to be imposed at the box walls. This field is an infinite array of oscillators with different momentum $\mathbf{k}$. To relate it to the quantum theory, we will define a real scalar field,

$$\varphi(x) = \zeta(x)\sqrt{\frac{\omega_0^3}{V}} = \sum_{\mathbf{k}}(2\omega V)^{-1/2}[\omega_0 T_{0\mathbf{k}} e^{-ikx} + \omega_0 T_{0\mathbf{k}}^* e^{ikx}]. \quad (55)$$

Following the same procedures as discussed in the previous section, we will quantize the field and promote $\zeta(x)$, $\varphi(x)$ and $T_{0\mathbf{k}}$ as operators. The creation and annihilation operators are defined as

$$a_{\mathbf{k}}^\dagger = \omega_0 T_{0\mathbf{k}}^\dagger, \quad (56)$$

$$a_{\mathbf{k}} = \omega_0 T_{0\mathbf{k}}. \quad (57)$$

These operators satisfy the commutation relations,

$$[a_{\mathbf{k}}, a_{\mathbf{k}'}^\dagger] = \delta_{\mathbf{k}\mathbf{k}'}, \quad (58)$$

$$[a_{\mathbf{k}}, a_{\mathbf{k}'}] = [a_{\mathbf{k}}^\dagger, a_{\mathbf{k}'}^\dagger] = 0, \quad (59)$$

$$[T_{0\mathbf{k}}, T_{0\mathbf{k}'}^\dagger] = \frac{\delta_{\mathbf{k}\mathbf{k}'}}{\omega_0^2}, \quad (60)$$

$$[T_{0\mathbf{k}}, T_{0\mathbf{k}'}] = [T_{0\mathbf{k}}^\dagger, T_{0\mathbf{k}'}^\dagger] = 0. \quad (61)$$

In terms of the creation and annihilation operators, equation (55) can be rewritten as

$$\varphi(x) = \sum_{\mathbf{k}}(2\omega V)^{-1/2}[a_{\mathbf{k}} e^{-ikx} + a_{\mathbf{k}}^\dagger e^{ikx}], \quad (62)$$

which is the bosonic field in quantum theory. Thus, particles are created and annihilated the same way as formulated in the standard quantum theory, except a particle observed also has oscillation in proper time.

By replacing the creation and annihilation operators with equations (56) and (57), we can rewrite some of the bosonic field's operators in terms of $T_{0\mathbf{k}}$ and $T_{0\mathbf{k}}^\dagger$. For example, the conjugate momenta of $\varphi(x)$, the particle number operator and the Hamiltonian of the system are,

$$\Pi(x) = \sum_{\mathbf{k}} -i\sqrt{\frac{\omega}{2V}}[\omega_0 T_{0\mathbf{k}} e^{-ikx} - \omega_0 T_{0\mathbf{k}}^* e^{ikx}], \tag{63}$$

$$N_{\mathbf{k}} = a_{\mathbf{k}}^\dagger a_{\mathbf{k}} = \omega_0^2 T_{0\mathbf{k}}^\dagger T_{0\mathbf{k}}, \tag{64}$$

and

$$H = \sum_{\mathbf{k}} \omega\left(a_{\mathbf{k}}^\dagger a_{\mathbf{k}} + \frac{1}{2}\right) = \sum_{\mathbf{k}} \omega\left(\omega_0^2 T_{0\mathbf{k}}^\dagger T_{0\mathbf{k}} + \frac{1}{2}\right), \tag{65}$$

where normal ordering shall be taken between $T_{0\mathbf{k}}$ and $T_{0\mathbf{k}}^\dagger$. Other examples are straightforward, which we will not repeat here. As we have illustrated, $\zeta(x)$ is a field with an infinite array of oscillators. A field with vibrations of matter in time and space can mimic the properties of a bosonic field.

Next, let us consider the Lagrangian density of $\zeta(x)$, i.e.,

$$\mathcal{L} = \frac{\bar{\rho}_m \omega_0^2}{2}[(\partial_0 \zeta)^2 - (\boldsymbol{\nabla}\zeta)^2 - \omega_0^2 \zeta^2], \tag{66}$$

where $\bar{\rho}_m = \omega_0/V$ is a mass density constant of the system. Based on this Lagrangian density, the conjugate momentum of $\zeta(x)$ is

$$\eta(x) = \frac{\partial \mathcal{L}}{\partial[\partial_0 \zeta(x)]} = \frac{-i\bar{\rho}_m \omega_0^2}{\sqrt{2}}\sum_{\mathbf{k}}[\tilde{T}_{\mathbf{k}} e^{-ikx} - \tilde{T}_{\mathbf{k}}^\dagger e^{ikx}]. \tag{67}$$

where

$$\tilde{T}_{\mathbf{k}} = \sqrt{\frac{\omega}{\omega_0}} T_{0\mathbf{k}} = \sqrt{\frac{\omega}{\omega_0^3}} a_{\mathbf{k}}, \tag{68}$$

$$\tilde{T}_{\mathbf{k}}^\dagger = \sqrt{\frac{\omega}{\omega_0}} T_{0\mathbf{k}}^\dagger = \sqrt{\frac{\omega}{\omega_0^3}} a_{\mathbf{k}}^\dagger. \tag{69}$$

The conjugate pair $\zeta(x)$ and $\eta(x)$ satisfy the equal-time commutation relations,

$$[\zeta(t,\mathbf{x}), \eta(t,\mathbf{x}')] = i\delta(\mathbf{x} - \mathbf{x}'), \tag{70}$$

$$[\zeta(t,\mathbf{x}), \zeta(t,\mathbf{x}')] = [\eta(t,\mathbf{x}), \eta(t,\mathbf{x}')] = 0. \tag{71}$$

The displaced time in $\zeta(x)$ is given by equation (12),

$$t_d(x) = \zeta_t(x) = \partial_0 \zeta(x) = \sum_{\mathbf{k}} \frac{-i}{\sqrt{2}}[\tilde{T}_{\mathbf{k}} e^{-ikx} - \tilde{T}_{\mathbf{k}}^\dagger e^{ikx}]. \tag{72}$$

Its rate relative to the external time $t$ is,

$$u_d(x) = \partial_0 t_d(x) = \sum_{\mathbf{k}} \frac{-\omega}{\sqrt{2}}[\tilde{T}_{\mathbf{k}} e^{-ikx} + \tilde{T}_{\mathbf{k}}^\dagger e^{ikx}]. \tag{73}$$

Comparing equations (67) and (72), we have

$$\eta(x) = \bar{\rho}_m \omega_0^2 t_d(x). \tag{74}$$

Therefore, $\zeta(x)$ and $t_d(x)$ can also form a conjugate pair, which satisfy equal-time commutation relations,

$$[\zeta(t,\mathbf{x}), t_d(t,\mathbf{x}')] = (\bar{\rho}_m \omega_0^2)^{-1}\delta(\mathbf{x} - \mathbf{x}'), \tag{75}$$

$$[t_d(t,\mathbf{x}), t_d(t,\mathbf{x}')] = 0. \tag{76}$$

From quantum theory, $\varphi(x)$ and its conjugate momenta are self-adjoint operators. This implies $\zeta(x)$, $\eta(x)$ and $t_d(x)$ are also the same. As we shall note, the vibrations of matter in time can include displacement in both the positive and negative directions. The spectrum of $t_d(x)$ can, therefore, span the whole real line[2].

From equation (5), the internal time of the matter field is

$$t_f(t,\mathbf{x}) = t + t_d(t,\mathbf{x}). \tag{77}$$

---

[2] For convenience purposes, we have chosen to study the conjugate pair formed by $\zeta(x)$ and $t_d(x)$, instead of using $u_d(x)$. Both options will lead us to the same conclusion that $t_d(x)$ does not form a conjugate pair with the Hamiltonian $H'$.

As discussed, the external time $t$ is a parameter. Since $t_d(t, \mathbf{x})$ is a self-adjoint operator, this implies that the internal time $t_f(t, \mathbf{x})$ can also be treated as a self-ajoint operator. Based on equations (75) and (76), $\zeta(x)$ and $t_f(t, \mathbf{x})$ shall satisfy the equal-time commutation relations,

$$[\zeta(t, \mathbf{x}), t_f(t, \mathbf{x}')] = (\bar{\rho}_m \omega_0^2)^{-1} \delta(\mathbf{x} - \mathbf{x}'), \tag{78}$$

$$[t_f(t, \mathbf{x}), t_f(t, \mathbf{x}')] = 0. \tag{79}$$

Since $t_f(t, \mathbf{x})$ does not form a conjugate pair with the Hamiltonian, this internal time operator can be treated as a self-adjoint operator without conflict with the Pauli's theorem.

The field $\zeta$ with vibrations of matter in space and time is directly related to the real scalar quantum field $\varphi$. As shown in equation (55),

$$\zeta(x) = \frac{\varphi(x)}{\omega_0 \sqrt{\bar{\rho}_m}}. \tag{80}$$

Similarly, by comparing equations (63) and (72), we have

$$t_d(x) = \frac{\Pi(x)}{\omega_0 \sqrt{\bar{\rho}_m}}. \tag{81}$$

Therefore, we can obtain the solutions for $\zeta$ and $t_d$ directly from those developed in the quantum field theory for $\varphi$ and $\Pi$ by multiplying a factor $(\omega_0 \sqrt{\bar{\rho}_m})^{-1}$. The field $\zeta$ not only allows the creation and annihilation of particles, but each particle also has proper time oscillation with specific amplitude $|\mathring{T}_0| = 1/\omega_0$. The displaced times of the oscillators are measured by the observable $t_d$.

## 8. Conclusions and discussions

The mathematics of a quantum field has its roots in the quantum harmonic oscillator. Instead of considering the raising and lowering of energy levels, we are dealing with the creation and annihilation of particles in a quantum field. Apart from that, the two systems' formulations have many similarities, which leads us to ponder whether there could be a deeper correlation between the two. In this paper, we demonstrate that we can define a proper time oscillating system in a similar way as we formulate a one-dimensional quantum harmonic oscillator. The concept can be applied to reconcile the properties of a zero-spin quantum field. While the particle in a quantum harmonic oscillator has oscillation in space, a proper time oscillator has oscillation in time. The introduced degree of freedom can allow a more symmetrical treatment between time and space in a quantized field.

Nature has a preference for symmetry [43]. If general relativity demands time and space to be treated on an equal footing, it is not implausible that time can have a more dynamic role in quantum theory. Not only the properties of a quantum field can be reconciled by allowing matter to vibrate in time, but we can also procure an internal time operator that is self-adjoint. The difficulties with Pauli's theorem can be overcome. With a self-adjoint time operator, the formulations will become more symmetrical, assuaging some of the differences between quantum theory and general relativity when we attempt to reconcile the two fundamental theories.

Heisenberg's uncertainty relation plays an essential role in the interpretation of the quantum theory. If time and space are on an equal footing, we will naturally expect time to satisfy an uncertainty principle with interpretation similar to the one for position. However, because of the absence of a self-adjoint time operator, it is generally not attainable. Unlike the standard approaches, we have demonstrated that a proper time oscillator's displaced internal time and displaced time rate satisfy an uncertainty relation that resembles the one between position and momentum. We do not need to confine our interpretation of time as a parameter since the internal time can be treated as a self-adjoint operator. Again, the adoption of such an uncertainty relation will permit a more symmetrical formulation for a quantized field.

In quantum theory, particles are treated as sets of coupled oscillators with their own de Broglie's internal clock. Interestingly, the proper time oscillators meet these criteria. If a particle has proper time oscillation, it should be possible to demonstrate these effects in an experiment. One possible way to observe the temporal oscillation is to measure the decay rate of an unstable particle, which will vary at different phases of the oscillation. Based on equation (35), the magnitude of a particle's oscillation amplitudes can be amplified and the frequency can be lowered by projecting the particle to a higher speed. Therefore, at a sufficiently high energy level, it is theoretically possible to slow down the oscillation of an unstable particle to detect a varying decay rate within one cycle of oscillation.

Another possible way to observe the effects of the temporal oscillation is to measure a neutrino's time of arrival. In recent years, many experiments have been conducted on the measurement of a neutrino's speed [44–50]. Although no deviation from the prediction of special relativity has been observed so far, experiments are continued for theoretical reasons, e.g possible existence of superluminal tachyons [51], Lorentz violating neutrino oscillation [52], lightcone fluctuation [53–55]. As proposed in the study of lightcone fluctuation, a neutrino can experience varying travel distance/time through fluctuating spacetime. This fluctuation can result in an accumulated uncertainty of a neutrino's travel distance/arrival time. Furthermore, it is suggested that these uncertainties are functions of the particle's energy [56–59], albeit the exact solutions for these functions are not yet known, and will have to be established either by experiments or theoretical predictions. Although the theory propounded in this paper is not about quantum spacetime, the results obtained share some similarities with those conjectured in the study of lightcone fluctuation.

As discussed in section 5, two neutrinos with the same average velocity can reach a target at different times, depending on the phases of their temporal oscillations. This deviation will result in an uncertainty of arrival time when we measure a large collection of particles. From equation (35), the oscillation amplitudes are directly proportional to the square root of the particle's energy. Therefore, the uncertainty of neutrinos' arrival time can also be magnified at a higher energy level, which can make the effect easier for detection. Because of their extreme light weight and our ability to project them to very high speed, neutrinos can be practically used for detecting the particles' oscillation effects.

## Data availability statement

All data that support the findings of this study are included within the article (and any supplementary files).

## ORCID iDs

Hou Y Yau  https://orcid.org/0000-0002-1938-3000